\documentclass[aps,pra,twocolumn,showpacs,showkeys,floatfix,superscriptaddress,10pt]{revtex4}
\usepackage[colorlinks,citecolor=blue,linkcolor=red,dvipdfm]{hyperref}
\usepackage{amsmath,amssymb,graphicx}
\usepackage{times}

\begin{document}

\title{Two-dimensional matter-wave solitons and vortices in competing cubic-quintic nonlinear lattices }
\author{Xuzhen Gao}
\affiliation{State Key Laboratory of Transient Optics and Photonics, Xi'an
Institute of Optics and Precision Mechanics of CAS, Xi'an 710119, China}
\affiliation{University of Chinese Academy of Sciences, Beijing 100084, China}

\author{Jianhua Zeng}
\email{\underline{zengjh@opt.ac.cn} or \underline{zengjianhua1981@gmail.com}}
\affiliation{State Key Laboratory of Transient Optics and Photonics, Xi'an
Institute of Optics and Precision Mechanics of CAS, Xi'an 710119, China}

\begin{abstract}
The nonlinear lattice---a new and nonlinear class of periodic potentials---was recently introduced to generate various nonlinear localized modes. Several attempts failed to stabilize two-dimensional (2D) solitons against their intrinsic critical collapse in Kerr media. Here, we provide a possibility for supporting 2D matter-wave solitons and vortices in an extended setting---the cubic and quintic model---by introducing another nonlinear lattice whose period is controllable and can be different from its cubic counterpart, to its quintic nonlinearity, therefore making a fully `nonlinear quasi-crystal'.

A variational approximation based on Gaussian ansatz is developed for the fundamental solitons and in particular, their stability exactly follows the inverted \textit{Vakhitov-Kolokolov} stability criterion, whereas the vortex solitons are only studied by means of numerical methods. Stability regions for two types of localized mode---the fundamental and vortex solitons---are provided. A noteworthy feature of the localized solutions is that the vortex solitons are stable only when the period of the quintic nonlinear lattice is the same as the cubic one or when the quintic nonlinearity is constant, while the stable fundamental solitons can be created under looser conditions. Our physical setting (cubic-quintic model) is in the framework of the Gross-Pitaevskii equation (GPE) or nonlinear Schr\"{o}dinger equation, the predicted localized modes thus may be implemented in Bose-Einstein condensates and nonlinear optical media with tunable cubic and quintic nonlinearities.
%\textbf{Keywords} {solitons and vortices, Bose-Einstein condensates, periodic potentials}

\end{abstract}



\maketitle

%\large
%\normalsize

\section{Introduction}

In recent years, the study of Bose-Einstein condensates (BECs) in ultracold alkali gases has attracted considerable theoretical and experimental interest \cite{BEC1,BEC2,BEC3,BEC4}. The significance of this field is multi-fold. It not only stimulated the development and application of cutting-edge techniques in condensed-matter physics, and nonlinear atomic and molecular physics for atomic cooling, trapping, and manipulations, but also provided a rich and interesting platform for theoretical predictions (both analytical and numerical), and a comparison with their experimental observations. BEC studies opened a research avenue on the condensation of bosonic atoms and offered new attention and in-depth analyses of other fascinating degenerate quantum gases, ranging from the condensation of Fermi gases \cite{Fermi} and Bose-Fermi mixtures to ultracold molecules \cite{Molecules,Molecules2}. They gave rise to many revolutionary implications, \emph{e.g.,} matter-wave interferometry, ultracold atomic clocks with unprecedented precision, and quantum information processing. Moreover, they shed new light on the investigation of collective phenomena related to those predicted many years ago, but never observed in many branches of physics (especially in solids, fluids, and nuclei). Among them are the quantum fluids and quantum-phase transitions \cite{BEC1,BEC2,BEC3,BEC4}.

Within a fundamental mean-field theory, the dynamics of matter waves in BECs can be described by the Gross-Pitaevskii equation (GPE) \cite{BEC1,BEC2}, which is generally called the nonlinear Schr\"{o}dinger equation (NLSE) as they are of the same form. Such theoretical models have had great success because the GPE can explain many experimental observations, e.g., the shape of the BECs, expansion, and collective excitations \cite{GP1,GP2,GP3}. The BEC is an intrinsic nonlinear media due to the existence of atom-atom collisions, which can be represented by nonlinear terms in the GPE \cite{BEC1,BEC2}. Many interesting nonlinear phenomena that arise in such ultracold degenerate quantum gases were predicted theoretically (by the GPE) and subsequently observed experimentally. These include the generation of dark solitons \cite{DS1,DS1b,DS1c,DS2,DS2b} (ref. a recent review \cite{DS3}), fundamental bright \cite{BS,BSb,BSc,BSd} and gap solitons \cite{GS}, vortices (solitons with embedded vorticity), and related structures \cite{vortex,vortex2,vortex3,vortex4,vortex5}, to name a few. It is commonly known that the balanced interplay of dispersion/diffraction and nonlinearity can create various localized modes, including dark and bright solitons, which exist under defocusing and self-focusing nonlinearities, respectively \cite{FPC}. By tuning the atom-atom interactions in BECs, the nonlinearity can be readily set to defocusing or self-focusing \cite{BEC1,BEC2}.

In the one-dimensional (1D) case, the GPE is an analytical solvable model and thus permits exact solutions for both dark and bright solitons \cite{BEC4}. In uniform media, the 1D bright solitons, supported by the cubic self-focusing nonlinearity, are exceptionally stable and robust. However, the stability of their multidimensional (two- and three-dimensional, 2D and 3D) counterparts is fragile, as the cubic self-focusing leads to the well-known phenomena of wave collapse (also known as ``blowing up'' in mathematical literature) or catastrophic self-focusing, making the fundamental multidimensional localized states unstable in free space; see \emph{e.g.,} some closely correlated reviews \cite{soliton-Rev1,soliton-Rev1b,soliton-Rev2,soliton-Rev2b,soliton-Rev3,soliton-Rev4,
soliton-Rev5,soliton-Rev5b} and the more comprehensive books \cite{soliton-book1,soliton-book2}.

It is well known that the 2D localized modes (\emph{viz.}, the \emph{Townes solitons} \cite{Townes}), supported by the cubic self-focusing nonlinearity, are restricted by the instability introduced by the critical collapse, and accordingly, have not yet been observed in experiments. Under the same background, more complex multidimensional localized states---vortical solitons (\emph{alias} vortex tori)---would again be subject to a still stronger azimuthal instability, making them split into fragments, which finally suffer an intrinsic collapse \cite{soliton-Rev1,soliton-Rev2}. It is, therefore, an issue of significant importance to the stabilization of multidimensional localized states, which, accordingly, has attracted much attention in the past years.

To this end, many approaches were tried to stabilize multidimensional fundamental localized
states---solitons and solitary vortices. A general method is to use trapping potentials; \emph{i.e.}, the external harmonic-oscillator trapping potentials \cite{HO,HO2}, which are axially and spherically symmetric, were utilized to trap, respectively, 2D and 3D fundamental matter-wave solitons and solitary vortices (with embedded vorticity S = 1), in self-attractive BECs. A more popular and promising method is the application of periodic potentials---optical lattices, which can be readily realized in experiments via the interference of multiple counter-propagating laser beams \cite{BEC1,BEC2,BEC3,BEC4,soliton-Rev2,soliton-Rev2b, Peli, BEC-Rev1, BEC-Rev2,NL, KVT}---rather than the normal harmonic-oscillator potentials.

Since the introduction of periodic structures, the creation of localized states \cite{BEC2,BEC4,soliton-Rev2, Peli, BEC-Rev1, BEC-Rev2,NL, KVT, BBM,YM} has been endowed with more prolific meanings; \emph{e.g.,} in addition to the fundamental solitons that exist under the self-focusing nonlinearity in the semi-infinite gap of the underlying linear spectrum, a type of gap soliton supported by defocused nonlinearity can also exist in the finite band gaps of the spectrum. The stabilization of multidimensional localized states (both 2D and 3D) were predicted in the forms of fundamental and gap solitons, as well as their vortical counterparts---solitary and gap vortices \cite{soliton-Rev2, Peli, BEC-Rev1, BEC-Rev2,NL, KVT, BBM,BBM2,YM}. Furthermore, the combination of a harmonically confining magnetic trap and an optical lattice was also used to create various localized modes \cite{HOOL}. It is necessary to point out that 1D gap solitons were realized experimentally in self-repulsive BECs \cite{GS}.

In addition to the atomic BECs, other physical realizations of multidimensional localized states, supported by periodic potentials, were extended to several optical structures, ranging from photonic crystals \cite{FPC, PC} to semiconductor microcavities and photorefractive optically induced photonic lattices (called photonic lattices, for simplicity, in many studies) \cite{PL1,PL2}. With the aid of spatially periodic potentials (alias lattice potentials), the stabilized mechanisms of multidimensional (both 2D and 3D) fundamental and vortical solitons were predicted in various nonlinear optics settings (mainly in photonic crystals and photonic lattices) \cite{FPC,PC,PL1,PL2}. Experimental realizations were, however, only demonstrated in the 2D cases, which included 2D optical vortex solitons in photonic lattices \cite{vortexPL,vortexPL2} and 2D plasmon-polariton gap solitons \cite{PPGS}, which are in the form of polariton condensates (alias exciton-polariton BECs) in semiconductor microcavities with a lattice structure.

The periodic potentials are recently extended to their nonlinear counterparts (sometimes called
pseudopotentials), alias \emph{nonlinear lattices} (NLs), which are characterized by nonlinear potentials with spatially periodic modulations of the sign and/or local strength of the nonlinearity (see a comprehensive investigation in a recent review \cite{NL} and references therein). The 1D NLs \cite{NL1D,NL1D2} and the combined linear-nonlinear lattices \cite{LNL1D, LNL2D} were widely used for studying various localized states. However, stabilizing 2D solitons against a critical collapse using purely NLs is still a challenging work. Although nonlinear shapes with sharp edges (\emph{e.g.}, circles or stripes) can support stable 2D solitons \cite{2DHS,YBVL, NL2D,NL2Db,NL2Dc}, which are essentially equal to the cases supported by a single circle, and the periodicity (of the NLs) does not play a dominant role in the stabilization \cite{YBVL}. With regard to this, a noteworthy work is on the experimental observation of NL-supported optical solitons formed at the interface between two lattices (solitons of this type are known as surface solitons) \cite{NLsee}.

This work is focused on a stabilized mechanism for the formation of 2D fundamental matter-wave and vortical solitons using purely NLs. Our physical setting is based on the commonly used cubic-quintic model by introducing NLs in both cubic and quintic nonlinearities, thereby forming a nonlinear \emph{quasi-lattice}. Although the study of soliton properties in the cubic-quintic model was widely reported \cite{CQ1,CQ1b,CQ2,CQ2b,CQ2c,CQ3,CQ3b,CQ3c}, the case for a spatially periodic modulation of both cubic and quintic nonlinear terms has not yet been investigated well (a recent work in \cite{CQ1D} confirmed that the 1D version of the cubic-quintic NLs can stabilize solitons against the critical collapse to some extent).

In the combined cubic and quintic NLs model, we show how 2D localized states can be created and stabilized via competing self-focusing cubic and self-defocusing quintic nonlinearities. The fundamental solitons are studied by means of numerical methods and a variational approximation. Merely numerical methods are used for vortex solitons, as taking an analytical approach is evidently an extremely difficult task. Stability regions for the fundamental and vortex solitons are identified. In particular, the Gaussian ansatz can match up with its numerical counterpart for fundamental solitons, whose stability condition is found to obey the inverted \textit{Vakhitov-Kolokolov} (anti-VK) stability criterion \cite{LNL1D, LNL2D,VK}, $d\mu/dN>0$, as shown below.

For physical realizations, the quintic nonlinearity in the GPE arises from three-body interactions in a dense BEC \cite{BEC1,BEC2,BEC3,BEC4}. In diverse nonlinear optical media \cite{CQ1D}, \emph{e.g.,} glass, liquids, and ferroelectric films, the quintic nonlinearity usually appears together with the cubic term. It should be noted that 2D spatial fundamental solitons in liquid carbon disulfide (a bulk optical media) with a competing cubic-quintic nonlinearity have recently been generated in experiments \cite{ECGHV}. Therefore, in addition to the matter waves in BECs, the theoretical results predicted here can also be realized in other physical settings---\emph{e.g.,} nonlinear optical media---by filling the holes of photonic crystals with index-matching materials such as liquids.

The paper is organized as follows. After introducing the theoretical model (GPE/NLSE) and its variational approximation grounded on the usual Gaussian ansatz in Sec. II, the numerical results for the relevant 2D localized modes---both fundamental and vortical solitons---and their stability regions, obtained by direct simulations of thus-found stationary solutions under weak perturbations, are presented in Sec. III. Finally, the paper is summarized in Sec. IV.

\section{Our model and its variational approximation}

\subsection{ Gross-Pitaevskii equation}

In terms of physical setting, our theoretical model is based on the normalized form of the underlying GPE (or NLSE) for the mean-field wave function (or the amplitude of an electromagnetic wave flowing in nonlinear optical media), $\psi (\mathbf{r},z)$:
\begin{eqnarray}
i\psi _{t}=-\frac{1}{2}(\partial _{x}^{2}\psi+\partial _{y}^{2}\psi)+\epsilon[\cos (2x)+\cos (2y)]|\psi |^{2}\psi \notag\\
 + g [\cos (qx)+\cos (qy)]|\psi |^{4}\psi,  \label{GPE}
\end{eqnarray}%
where time parameter $t$ is replaced by propagation distance $z$ for electromagnetic-wave propagation in nonlinear optics; $\varepsilon$ and $g$, separately, are the strengths of the cubic-quintic NLs. At $g=0$, Eq. (\ref{GPE}) is equal to the cubic nonlinear model with the spatially periodic nonlinearity, \emph{i.e.,} NL, where 2D solitons cannot be stabilized at all. Considering that the center of the soliton would be placed at point $x=y=0$, $\varepsilon <0$ corresponds to the self-focusing cubic nonlinearity, with $g>0$ being the defocusing quintic term. Throughout this paper, unless otherwise specifically mentioned, we set $\varepsilon \equiv -3$.

The remaining variable $q$ in the quintic NL will be given below. It is worth mentioning that the cubic NL and quintic NL are commensurate at $q = 2$; the subharmonic commensurability appears at $q = 1$, while it is incommensurate when $q$ is at other values. Therefore, the cubic NL and quintic NL are in different periods (spatial arrangements) with the variation of $q$, making the medium virtually equivalent to a purely nonlinear \emph{quasi-lattice} (a \emph{quasi-crystal} for nonlinear excitations). Obviously, the quintic nonlinearity is uniform at $q = 0$.

Stationary solutions to Eq. (\ref{GPE}) with chemical potential $\mu $ (or
propagation constant $-\mu $, for nonlinear optical waves) are sought in the form of
$\psi (x,y,t)=\phi (x,y)\exp (-i\mu t)$, with wave function $\phi(x,y)$
yielding the following stationary equation:
\begin{eqnarray}
\mu \phi=-\frac{1}{2}(\partial _{x}^{2}\phi+\partial _{y}^{2}\phi)+\epsilon[\cos (2x)+\cos (2y)]|\phi |^{2}\phi \notag\\
 + g [\cos (qx)+\cos (qy)]|\phi |^{4}\phi .  \label{phi}
\end{eqnarray}%
The above equation can be directly derived from its Lagrangian form, \newline
\begin{eqnarray}
L=\frac{1}{2}\int_{-\infty}^{+\infty}\{\mu |\phi |^{2}-\frac{1}{2}(|\frac{\partial \phi}{\partial x}
|^{2}+|\frac{\partial \phi}{\partial y}|^{2})-\frac{\epsilon}{2}[\cos(2x)+\notag\\
\cos(2y)]|\phi|^{4}-\frac{g}{3}[\cos(qx)+\cos(qy)]|\phi|^{6}\}dxdy .   \notag\\
\label{Lagran}
\end{eqnarray}

\subsection{Variational approximation}

Variational approaches are generally used to study stationary solutions, particularly for fundamental modes, since they can predict the shape and even the stability condition (used in combination with the stability criterion of Ref. \cite{VK}), which can supplement, perfect, and verify the results obtained by direct simulations. To implement a variational approximation, we take a Gaussian ansatz as usual, $\phi \left( x,y\right) =A\exp \left[ -\left(
x^{2}+y^{2}\right) /\left( 2W^{2}\right) \right] $, with width $W$; the corresponding
norm (alias number of atoms) $N\equiv \int \int \phi ^{2}\left( x,y\right) dxdy=\pi \left( AW\right)
^{2}$ (or the total power, in terms of optics). The substitution of such ansatz into Lagrangian Eq. (\ref{Lagran}), after simplification, leads to the following expression, written with variables $N$ and $W$:
\begin{equation}
L_{\mathrm{eff}}=\frac{N}{2}\left[ \mu -\frac{1}{2W^{2}}-\frac{\varepsilon N}{2\pi W^{2}}
e^{-\frac{W^2}{2}}-\frac{2gN^2}{9\pi^2 W^{4}}e^{-\frac{q^2W^2}{12}}\right],
\end{equation}%
and the corresponding variational equations, $\partial L_{\mathrm{eff}%
}/\partial N=\partial L_{\mathrm{eff}}/\partial W=0$:%
\begin{eqnarray}
\frac{\varepsilon N}{2\pi}(1+W^2)e^{-\frac{W^2}{2}}+
\frac{8gN^2}{9\pi^2 W^{2}}(1+ \frac{q^2 W^2}{24})e^{-\frac{q^2W^2}{12}} &=&-1 ,  \notag\\
\frac{1}{2W^{2}}+\frac{\varepsilon N}{\pi W^{2}}e^{-\frac{W^2}{2}}+
\frac{2gN^2}{3\pi^2 W^{4}}e^{-\frac{q^2W^2}{12}} &=&\mu .  \notag\\
\label{VA}
\end{eqnarray}%
In the following, the variational equations (\ref{VA}) based on the Gaussian ansatz will be solved numerically for fundamental soliton studies.

\begin{figure}[tbp]
\begin{center}
\includegraphics[width=1.0\columnwidth]{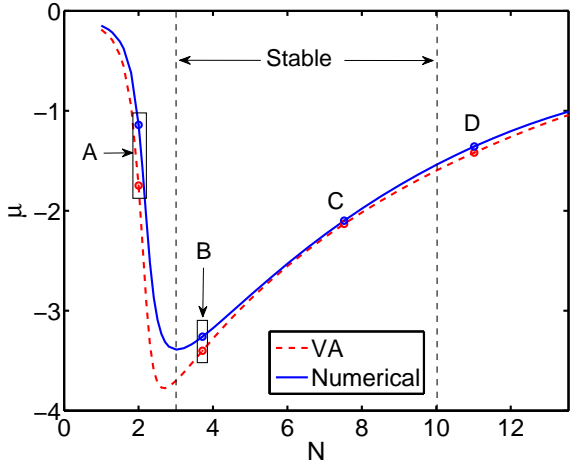}
\end{center}
\caption{ Chemical potential ($\mu$) vs. the norm ($N$) for 2D fundamental solitons
in the model with combined cubic and quintic periodic potentials (nonlinear lattices), produced by the variational approximation (VA, red curve), and found from numerical solutions of Eq. (\ref{phi}) (blue curve), at $\varepsilon =-3$, $g=1$, and $q=2$. Stable sections of the soliton families are within the marked stripe. Typical examples of stable and unstable fundamental solitons, corresponding to the marked points (B, C and A, D) are respectively displayed in Figs. (\ref{fig2}) and (\ref{fig3}).}
\label{fig1}
\end{figure}

\section{Numerical results for two-dimensional localized states}

Before proceeding with the numerical computations, we introduce our numerical methods. Specifically, the localized stationary modes (both fundamental and vortex solitons) were constructed numerically by means of the imaginary time-integration method \cite{FDTD} applied to Eq. (\ref{GPE}). The stability of the stationary solutions thus found against small perturbations was valuated through direct simulations of Eq. (\ref{GPE}) (in real time) using the finite-difference time-domain method \cite{FDTD}. It is relevant to stress that the localized stationary modes can also be found as numerical solutions of stationary equation (\ref{phi}) using Newton{'}s method. The numerical calculations were performed in a $30\times30$ domain on a grid of $256\times256$ points.

\subsection{Fundamental solitons}
We begin with the full commensurability case between the self-attractive cubic and self-repulsive quintic NLs (with the same period $\pi$), which is to say $q=2$ for the quintic term in Eq. (\ref{GPE}) and all the others.

\begin{figure}[tbp]
\begin{center}
\includegraphics[width=1.05\columnwidth]{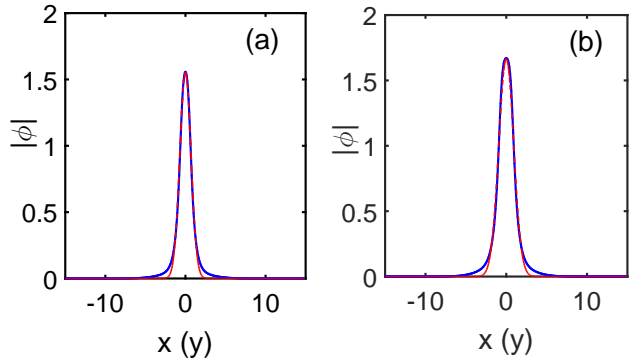}
\end{center}
\caption{ Typical examples of stable 2D fundamental solitons
found in the model with combined cubic and quintic nonlinear lattice potentials, at $\varepsilon =-3$, $g=1$, and $q=2$. Only contour plots of the modes (the modules of the stationary wave functions) are shown by projecting them onto a 2D plane. The left (a) and right (b) panels correspond, respectively, to the marked points B and C in Fig. (\ref{fig1}). Hereinafter, the blue solid and red-dashed curves are numerical stationary solutions and their relevant Gaussian ansatz, respectively. }
\label{fig2}
\end{figure}

\begin{figure}[tbp]
\begin{center}
\includegraphics[width=1.1\columnwidth]{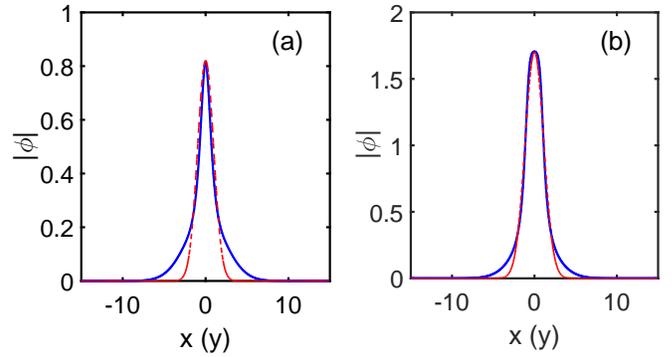}
\end{center}
\caption{ Typical examples of unstable 2D fundamental solitons
supported by a model with combined cubic and quintic nonlinear lattice potentials, at $\varepsilon =-3$, $g=1$, and $q=2$. The contour plots shown here are only the modes of the stationary wave functions in the 2D plane. The left (a) and right (b) panels correspond, respectively, to the points A and D in Fig. (\ref{fig1}). }
\label{fig3}
\end{figure}
\begin{figure}[tbp]
\begin{center}
\includegraphics[width=1.05\columnwidth]{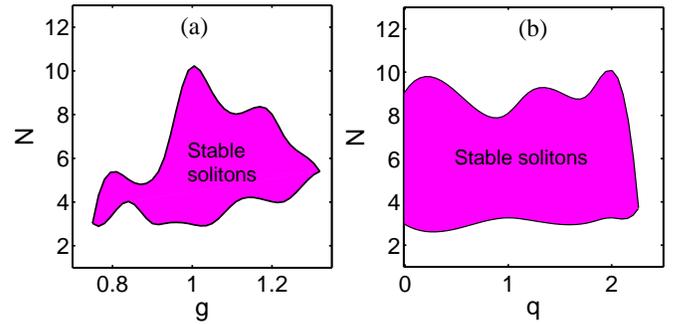}
\end{center}
\caption{ Stability borders for the entire set of 2D fundamental solitons
supported by cubic and quintic nonlinear lattice potentials: (a) curve $N(g)$ with different values of quintic nonlinear strength $g$ at $\varepsilon =-3$ and $q=2$; (b) curve $N(q)$ with different values of quintic nonlinear lattice structure variable $q$ at $\varepsilon =-3$ and $g=1$. Portions of the stable solitons are confined to the respective stability borders (the magenta areas). }
\label{fig4}
\end{figure}
Figure \ref{fig1} depicts the relations $\mu (N)$ for the fundamental solitons based on the variational approximation, produced by a numerical solution of variational equations (\ref{VA}). They were combined with their fully numerical counterparts produced by numerical stationary solutions of the stationary equation (\ref{phi}), and checked through direct simulations of the perturbed solutions in nonlinear evolution equation (\ref{GPE}), at $\varepsilon =-3$, $g=1$, and $q=2$.

Examples of stable 2D fundamental solitons, supported by the competing cubic-quintic nonlinear lattices, are shown in Fig. \ref{fig2}. These stable localized modes can match well with their Gaussian ansatz, and are quasi-isotropic and highly localized within a single cell (recall that the period here is $\pi$, for both cubic and quintic NLs). The former feature is natural as our system meets spatial-inversion symmetry: \emph{e.g.,} equation (\ref{GPE}) or equation (\ref{phi}) is invariant under the symmetry operations $x \rightarrow -x$ and $y \rightarrow -y$. The latter case may be explained by the fact that, to arrest a 2D critical collapse, the localized modes should reside themselves into a single well.

However, as seen from Fig. \ref{fig3}, although broad solitons---their widths are much bigger than the NLs¡¯ relevant periodicity $\pi$ (recall that both the cubic and quintic NLs have the same period at $q=2$)--- also exist, direct simulations verified that they are totally unstable, in contrast to their 1D counterparts, where the broad solitons and solitons with symmetric side peaks are found to be stable; cf. Figs. 5 and 7 in ref. \cite{CQ1D}.

\begin{figure}[tbp]
\begin{center}
\includegraphics[width=1\columnwidth]{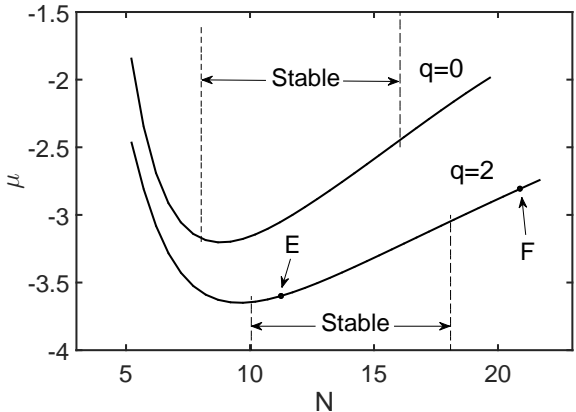}
\end{center}
\caption{Numerically found curves $\mu(N)$ for families of 2D vortex solitons with topological charge 1, supported by the cubic and quintic nonlinear lattice potentials ($q=2$) and constant quintic term ($q=0$), at $\varepsilon =-3$ and $g=1$. The stable vortex solitons are limited to the corresponding marked stripes. Typical examples of the stable and unstable vortex solitons, marked by points E and F at $q =2$, are respectively depicted in Fig. \ref{fig6}.}
\label{fig5}
\end{figure}
\begin{figure}[tbp]
\begin{center}
\includegraphics[width=1.1\columnwidth]{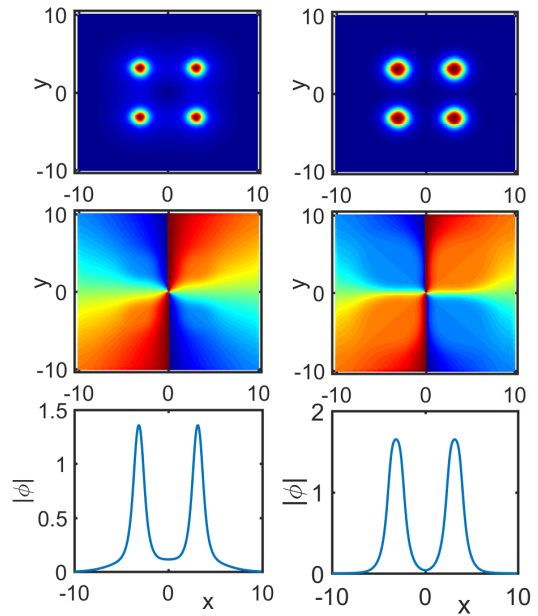}
\end{center}
\caption{Examples of stable (left panels) and unstable (right panels) 2D vortex solitons with topological charge 1, in the competing focusing cubic and defocusing quintic nonlinear lattices. The first, second, and third rows represent, respectively, the contour plots of wave function $|\phi|$, the phase distribution that carries the vorticity, and their cross-section profiles at $y=\pi$. The stable and unstable vortex solitons correspond respectively to points E and F in Fig. (\ref{fig5}).}
\label{fig6}
\end{figure}

From Fig. \ref{fig1}, we can see that the variational approximation provides a relatively reasonable accuracy for 2D fundamental solitons. Direct simulations of the perturbed solutions further demonstrated that they are stable only roughly around the norm $N\in [3,10]$, and their stability follows the ¡ùanti-VK¡ì criterion \cite{LNL1D, LNL2D,VK}, \emph{viz.,} $d\mu/dN>0$. The shapes of unstable 2D fundamental solitons (we call them broad solitons) in Fig. \ref{fig3} cover several lattice cells, making them unstable after some finite evolution time.

A vast number of direct simulations, made by changing variable $q$ and nonlinear coefficient $g$, demonstrate that the stable 2D fundamental solitons are only limited to the norm $N\in [N_L,N_H]$ (ref. the stable region marked in Fig. \ref{fig1}). This can be understood since, at the lower limit $N_L$, the stationary solutions are too weak to form stable localized modes, while above the threshold---the upper limit $N_H$---the stationary solutions are naturally unstable because of the critical collapse.

Fig. \ref{fig4}(a) summarizes the stability of the 2D fundamental solitons, supported by the combined cubic and quintic NLs, for various values of the quintic coefficient $g$. It is observed that, for the given cubic nonlinear strength $\varepsilon =-3$, the range of $g$ is within $[0.75,1.33]$ when allowed by the stable 2D fundamental solitons. This is natural since the cubic-focusing nonlinearity is overwhelming when its quintic-defocusing term $g$ is small; only above some certain value can the quintic nonlinearity help to arrest the critical collapse. The quintic coefficient $g$, however, cannot be too large either; otherwise, the quintic defocusing nonlinearity dominates, finally delocalizing the solitons.

Fig. \ref{fig4}(b) shows the stability region for the 2D fundamental solitons under various conditions, including the aforementioned incommensurate, subcommensurate, and commensurate relations between the cubic NL and quintic NL. The fundamental solitons are stable at $q <q _{\max}\approx 2.22$, for moderate values of $N$. When above such a threshold ($q _{\max}$), the incommensurability of the competing cubic-quintic NLs expands quickly, allowing the quintic NL to play a inappreciable role in stabilizing the localized modes, and in overcoming the critical collapse rooted in the cubic-focusing nonlinearity.

\subsection{Vortex solitons}

In our analysis, the ansatz for vortex solitons is taken as $\phi \left( x,y\right) =A\exp \left[ -\left(x^{2}+y^{2}\right) /\left( 2W^{2}\right) \right] e^{iS\theta(x,y)} $, with width $W$, topological charge $S$, and azimuthal coordinate $\theta$. Fig. \ref{fig5} displays the numerically found relations $\mu(N)$ for general 2D vortex solitons with topological charge 1, supported by the combined cubic and quintic NLs, at $\varepsilon =-3$ and $g=1$. Both the constant quintic nonlinearity ($q=0$) and quintic NL ($q=2$, commensurate cubic-quintic NLs) are shown in the figure. Direct simulations demonstrated that the stable 2D vortex solitons are within limited scopes and at moderate norm $N$.

The left column of Fig. \ref{fig6} shows a typical example of a stable vortex soliton with topological charge 1, which was created as hollow four-peak complexes, with a separation between the peaks of two times the period of the NL potentials, and an empty site at the center. As reported previously in other periodic potentials \cite{soliton-Rev2,NL,LNL2D}, such vortex types with inner voids can be stable because of the weak interaction between the peaks. Such a hollow vortex structure was constructed, rather than the normal isotropic vortex that resides within a single cell of the nonlinear lattice, since our detailed numerical examination found that the latter one cannot exist in the current model.

As can be seen from the right column of Fig. \ref{fig6}, similar to their unstable fundamental counterparts in Fig. \ref{fig3}, the stationary solutions of unstable vortex solitons occupy more than a single well, which makes them hard to station there. It is relevant to note that the wave structures of the stable and unstable vortex solitons at $q=0$ are very similar to their counterparts at $q=2$, as displayed in Fig. \ref{fig6}.

\begin{figure}[tbp]
\begin{center}
\includegraphics[width=1\columnwidth]{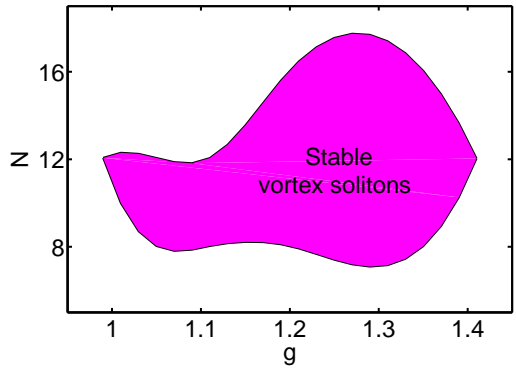}
\end{center}
\caption{ Stability border, shown as the curve $N(q)$, for the families of 2D vortex solitons, supported by the cubic and quintic nonlinear lattice potentials with $\varepsilon =-3$ and $q=2$, and under the different fifth-order (quintic) nonlinearity $g$. The vortex solitons are stable between the relevant stability borders (the magenta areas).}
\label{fig7}
\end{figure}

We obtained the stability areas of such hollow four-peak complexes (vortex soliton at topological charge 1) for different values of fifth-order nonlinear coefficient $g$ of the model through mass numerical simulations, which are shown in Fig. \ref{fig7}. Our calculations verified that the formation condition of stable vortex solitons is strictly under the full commensurability of the combined cubic NL and quintic NL ($q=2$) or under the model with constant quintic nonlinearity ($q=0$). Any case deviating from these two scenarios would be unstable; \emph{i.e.}, neither $q=0.1$ nor $q=1.9$ can generate stable vortex solitons.

\section{Conclusions}

We explored the 2D matter-wave fundamental and vortex solitons in the general cubic-quintic model, which is the Gross-Pitaevskii equation or nonlinear Schr\"{o}dinger equation, by introducing nonlinear lattices with tunable periods (incommensurate, subcommensurate, and commensurate) to both nonlinear terms, which can be recognized as a fully nonlinear quasi-crystal. The physical setting is the dense Bose-Einstein condensates under both two- and three-body interactions, with periodic changes of the scattering lengths of interatomic collisions by means of the Feshbach resonance. The cubic NL of this model is fixed, while both the strength and period of the quintic NL are variable.

A variational approximation was developed for the fundamental solitons and they were found to obey the anti-VK stability criterion, while the vortex solitons were studied by simply relying on numerical methods. Stability regions for both localized modes---fundamental and vortex solitons---were identified. In particular, stable vortex solitons existed as long as the quintic NL had the same period as its cubic term or the quintic nonlinearity was constant; in contrast, the formation conditions for the fundamental solitons were considerably more relaxed.

The physical model considered here can be extended to a two-component model---the coupled Gross-Pitaevskii equations \cite{CQ2}---to consider the existence of multidimensional localized modes therein. Considering that the experimental observation of 2D optical solitons in a cubic-quintic-septimal media was reported very recently \cite{QS}, it would be interesting, at least on a theoretical level, to add nonlinear lattices to these combined cubic-quintic-septimal nonlinearities (to one, two, or three nonlinear terms) and study the possible localized modes \cite{CQS}.

\section{Acknowledgments}

This work was supported by the NSFC, China (project Nos. 61690224, 61690222, 11204151), by the Youth Innovation Promotion Association of the Chinese Academy of Sciences (project No. 2016357) and the CAS/SAFEA International Partnership Program for Creative Research Teams, and partially by the Initiative Scientific Research Program of the State Key Laboratory of Transient Optics and Photonics.

\end{document}